\documentstyle[11pt]{article}

\begin{document}

%--------------------------------------------------------------
\begin{flushright}
	   KAIST-CHEP-95/20        \\
\end{flushright}
\begin{center}
   \Large{ The  Entropy of the  Complex Scalar Field    \\
       in a Charged Kerr Black Hole}
\end{center}
 
\vspace{2cm}
   
\begin{center}
 {Min-Ho Lee\footnote{e-mail~:~mhlee@chep6.kaist.ac.kr} and Jae 
 Kwan Kim     \\
 Department of Physics, Korea Advanced Institute of 
 Science and Technology \\
 373-1 Kusung-dong, Yusung-ku, Taejon 305-701, Korea.}
\end{center}

\vspace{2cm}
		 
\begin{abstract}
By using the brick wall method we calculate  the thermodynamic 
potential of the complex scalar field 
in a charged Kerr black hole.
Using it we show that 
in  the Hartle-Hawking state  
the leading term of the entropy is proportional to 
$   \frac{ A _H}{\epsilon^2}$, 
which  becomes divergent as the system approaches the 
black hole horizon.  
The  origin of the divergence is that the density of states
 diverges  at the horizon.
\end{abstract}
\vspace{1cm}
PACS numbers: 04.62.+v, 04.70.Dy

%---------------------------------------------------------------------
\newpage

  By  comparing  the black hole physics with the thermodynamics  
  Bekenstein  showed that the black hole entropy is proportional to the
  horizon area \cite{Bekenstein1,Hawking},
  and Hawking's discovery of  the black hole evaporation 
  confirmed that.
In   Euclidean path integral approach  it was shown that the tree level 
contribution of the gravitation action gives the black hole  
entropy \cite{Gibbons}.
However the exact statistical origin of the Bekenstein-Hawking 
black hole entropy is unclear.

Recently many efforts have been concentrated  on  understanding   
the statistical origin of  black hole thermodynamics, specially the 
black hole entropy by various methods \cite{Bekenstein2}: 
the brick wall method, the conical singularity
method, and the entanglement entropy method \cite{tHooft,Pad}. 
The leading term of the entropy obtained by those methods 
is proportional to the surface area of the  horizon.
However  the proportional coefficient diverges as the cut-off goes to 
zero.
The divergence is because of    an 
infinite number of states near the horizon,
which can be explained by the equivalence principle \cite{Barbon}.

 For the  black hole with a rotation  the entropy of the neutral scalar 
 field was calculated by authors \cite{MinHo}.
 It was shown that the leading term of the entropy, if the quantum 
 field is at  the Hartle-Hawking state, 
 is proportional to the horizon area.
 In this paper to understand more deeply the black hole entropy  
we will investigate   the entropy of the complex scalar field 
interacting with the gauge field $A_\mu$ in 
 the charged Kerr  black hole background \cite{Newman}.

%==============================================

Let us consider a minimally coupled complex scalar field with 
mass $\mu$ in thermal equilibrium at  temperature $1/\beta$  in the 
charged Kerr  black hole spacetime. 
 The line element of the charged Kerr black hole spacetime
and the electromagnetic vector  potential
in Boyer-Lindquist coordinates 
are given by \cite{Newman,Wald} 
\begin{eqnarray}
\nonumber
ds^2 &= & - \left( 
		\frac{ \Delta - a^2 \sin^2 \theta }{\Sigma}
		\right) 
		dt^2 - \frac{2 a \sin^2 \theta ~( r^2 + a^2 - \Delta)}{
		\Sigma } dt d\phi \\
\nonumber
    & &~ + \left[ \frac{(r^2 +a^2 )^2 - \Delta a^2 \sin^2 \theta }{
 \Sigma}  \right] \sin^2 \theta d \phi^2 + \frac{\Sigma}{\Delta} dr^2 +
    \Sigma ~d \theta^2 \\
&\equiv& g_{tt}(r, \theta) d t^2  
    + 2 g_{t \phi}( r,\theta) dt d \phi 
    +  g_{\phi \phi}(r, \theta) d\phi^2 
    + g_{rr} (r,\theta) d r^2 
    +  g_{\theta \theta}(r, \theta)   d\theta^2, 
               \label{metric}  \\
\nonumber
A_\mu &=& - \frac{e r}{\Sigma} \left[ (dt)_\mu 
	   - a \sin^2 \theta (d \phi)_\mu \right] \\
 &\equiv&  A_t(r, \theta) (d t)_\mu  + A_\phi (r,\theta) (d \phi)_\mu,
\end{eqnarray} 
where 
\begin{equation}
\Sigma = r^2 + a^2 \cos^2 \theta,~~~~~ \Delta = r^2 + a^2 + e^2 - 2 M r,
\end{equation}
and $e,a,$ and $M$ are charge, angular momentum per unit mass,
and mass of the spacetime respectively.
The charged Kerr black hole  spacetime  has two Killing vector
fields: the time-like Killing vector $\xi^\mu = (\partial_t)^\mu$
and the axial Killing vector $\psi^\mu = ( \partial_\phi)^\mu$.
We assume that $e^2 + a^2 < M^2$. In that case the  
 charged Kerr black hole has the  event horizon  
at $ r = r_H = M + \sqrt{M^2 - a^2 -e^2}$ 
and the stationary limit surface   at 
$r = r_0 = M + \sqrt{M^2 - e^2 - a^2 \cos^2 \theta }$. 

The equation of motion of   the  field  with mass $\mu$ is
\begin{equation}
\left[ (\nabla_\mu - i q A_\mu )( \nabla^\mu - 
i q A^\mu )  - \mu^2 \right] 
\Psi = 0.           \label{equation}
\end{equation}
We assume that the scalar field  is rotating with a constant 
azimuthal angular velocity $\Omega_0$ and has a chemical 
potential $\Phi_0$.  The  associated  conserved quantities are
the angular momentum $J$ and the electric charge $Q$.
For such a equilibrium ensemble of the states  of  the  field 
the grand partition function is given by 
\begin{equation}
Z =  {\rm Tr}  e^{-  \beta (H  - \Omega_0\cdot J - \Phi_0 Q  ) }.
\end{equation}
The thermodynamic potential of the system for particles with charge $q$ 
(for anti-particle  $q \rightarrow - q$) is given by 
\begin{equation}
 W  =  \frac{1}{\beta} \sum_{j,m} d_{j,m} \ln 
       \left( 1 - e^{- \beta( {\cal E}_{j,m} - m \Omega_0  
       - q \Phi_0)}  
       \right)  
       \label{po}
\end{equation}
or
\begin{equation}
    W =   \frac{1}{\beta} \sum_m \int_0^\infty d {\cal E}  
    g({\cal E}, m) \ln 
       \left( 1 - e^{- \beta( {\cal E} - m \Omega_0  - q \Phi_0)} 
       \right), 
\end{equation}
where $g({\cal E},m)$ is the density of states 
for a given ${\cal E}$ and $m$.

To evaluate the thermodynamic potential we will follow 
the brick wall method  of 't Hooft \cite{tHooft}.
Following the brick wall method we impose a small cut-off $h$
such that 
\begin{equation}
\Psi  (x) = 0 ~~~~{\rm for  }~~~ r \leq r_H + h.
\end{equation}
To remove the infra-red divergence   we also introduce another 
cut-off  $ L \gg r_H$ such that 
\begin{equation}
\Psi (x) = 0~~~~ {\rm for} ~~~r \geq L.
\end{equation}
In the WKB approximation with 
$\Psi = e^{- i {\cal E} t + i m \phi + i S(r, \theta)}$
the  equation (\ref{equation})  in Lorentz gauge  
$\nabla_\mu A^\mu = 0$  yields 
the constraint \cite{Mann}
\begin{equation}
  p_r^2 = \frac{1}{g^{rr}}  \left[
   - g^{tt} ({\cal E} + q A_t)^2 + 2 g^{t \phi} ({\cal E} 
   + q A_t) ( m  - q A_\phi)
       - g^{ \phi \phi } (m - q A_\phi)^2 
       - g^{ \theta \theta } p_\theta^2 
       - \mu^2      \right],         \label{con1}
\end{equation}
where  $  p_r = \partial_r S$ and $ p_\theta = \partial_\theta S$.
The number of states for a given  ${\cal E}$ is determined by  
$p_\theta, p_r$ and $m$.
Therefore the thermodynamic potential ( as in flat spacetime)
is written as
\begin{equation}
\beta W  = \int dr d \theta d\phi \sum_m \int 
\frac{d p_r  d p_\theta }{(2 \pi)^2} \ln  \left(
 1  - e^{ - \beta ( {\cal E} - m \Omega_0 - q \Phi_0 ) }   \right),
\end{equation}
where 
\begin{equation}
{\cal E} = \frac{1}{ - g^{tt}} 
      \left\{ - g^{t \phi}  p_\phi   +  
	 \left[
           \left(g^{t \phi} p_\phi  \right)^2 + ( - g^{tt}) 
	    \left( \mu^2 + 
    g^{\phi \phi} p_\phi^2 + g^{rr} p_r^2 + g^{\theta \theta}
	    p_\theta^2 \right)
         \right]^{1/2} 
     \right\}  - q A_t. 
\end{equation}
Here $p_\phi = m - q A_\phi$.
Let us introduce new variables:
\begin{equation}
\bar{p}_\phi =  \left(  \frac{ - {\cal D} }{g^2_{\phi \phi}}
		 \right)^{1/2} p_\phi,~~~
\bar{p}_r = \left( \frac{ - {\cal D}}{g_{\phi \phi} g_{rr} }
		 \right)^{1/2} p_r, ~~~
\bar{p}_\theta = \left( \frac{ - {\cal D}}{g_{\phi \phi} 
g_{\theta \theta} }
		 \right)^{1/2} p_\theta, 
\end{equation}
where ${\cal D} = g_{tt} g_{\phi \phi} - g_{t \phi}^2 $.
Then the thermodynamic potential,  with the assumption that $m$ is a 
continuous variable, is written as
\begin{eqnarray}
\nonumber
& &  \beta W \\
\nonumber
 &=& \int dr d \theta d\phi \int 
\frac{d \bar{p}_r  d \bar{p}_\theta  d \bar{p}_\phi}{(2 \pi)^3} 
\frac{g_{\phi \phi}^2 \sqrt{g_{\theta \theta} g_{rr}}
        }{ ( - {\cal D} )^{3/2}} 
\ln  \left(
 1  - e^{ - \beta (  \alpha \bar{p}_\phi +  
 \sqrt{ \bar{p}_\phi^2 + \bar{p}_r^2 
 + \bar{p}_\theta^2  + \bar{\mu}^2 }  - 
 q [ A_t + \Omega_0 A_\phi] - q \Phi_0 )  }   \right)           \\
 &=& \int dr d \theta d\phi \int 
\frac{d \bar{p} \bar{p}^2  d \bar{\theta} \sin \bar{\theta} 
d \bar{\phi}   }{ (2 \pi)^3 } 
\frac{g_{\phi \phi}^2 \sqrt{g_{\theta \theta} g_{rr}}
          }{ ( - {\cal D} )^{3/2}} 
\ln  \left(
 1  - e^{ - \beta (  \alpha \bar{p} \cos \bar{\theta} +  
 \sqrt{ \bar{p}^2 +  \bar{\mu}^2 }  - 
 q \Gamma )  }   \right)
\end{eqnarray}
in spherical coordinates in momentum space, 
where
\begin{equation}
\alpha =  \left( - \frac{g_{t \phi}}{g_{\phi \phi}}  - \Omega_0 \right)
	  \left( \frac{g_{\phi \phi}^2}{- {\cal D}} \right)^{1/2},~~~~~
  \bar{\mu}^2 = \left( \frac{ -{\cal D}}{ g_{\phi \phi}} \right) \mu^2,
	 ~~~~~\Gamma = \Phi_0 + A_t + \Omega_0 A_\phi.
\end{equation}

Note that  we must  restrict the system to be in the region such that 
 $  1 \pm \alpha  > 0$ or equivalently  $g_{tt}^{'} \equiv 
 g_{tt} + 2 \Omega_0 g_{t \phi} + \Omega_0^2 g_{\phi \phi} < 0$.
 In the region such that $ - g_{tt}^{'} >0$ ( called region I) 
 the integral is convergent.  But in the region 
 such that  $- g_{tt}^{'} \leq 0$
( called region II)  the integral is divergent.
 These facts  become more apparent if we investigate  
 the momentum phase space.  In the region  I 
the possible points of $p_i$  satisfying 
$ {\cal E}  + q A_t - \Omega_0 p_\phi  = E$ 
for a given $E$ are located on the following surface
\begin{equation}
\frac{p_r^2}{g_{rr}} + \frac{p_\theta^2}{g_{\theta \theta}} +
      \frac{- {g'}_{tt}}{- \cal D} \left(
              p_\phi + \frac{g_{t \phi }  
           + \Omega_0 g_{\phi \phi}}{{g'}_{tt}} 
           E  \right)^2 
   = \left( \frac{ E^2}{- {g'}_{tt} } 
   - \mu^2 \right),   \label{ellipsoid}
\end{equation}
which is a ellipsoid,  {\it a compact surface}. 
So the density of states $g(E)$ for a given $E$ is 
finite and the integrations 
over $p_i$  give a  finite value.
But in the region  II 
the possible points of $p_i$  are located on the following surface
\begin{equation}
\frac{p_r^2}{g_{rr}} + \frac{p_\theta^2}{g_{\theta \theta}} -
        \frac{ {g'}_{tt}}{- \cal D} \left( 
         p_\phi + \frac{g_{t \phi }  + 
         \Omega_0 g_{\phi \phi}}{{g'}_{tt}} 
	 E     \right)^2 
   =  - \left( 
             \frac{ E^2  }{  {g'}_{tt}} 
             + \mu^2 \right),
\end{equation}
which is  a hyperboloid,   {\it a non-compact surface}. So 
$g(E)$ diverges and the integrations over $p_i$  diverges. 
In case of $g^{'}_{tt} = 0$, the possible points   are 
given by the surface
\begin{equation}
\frac{p_r^2}{g_{rr}} + \frac{p_\theta^2}{g_{\theta \theta}} =
\frac{p_\phi - \left( \frac{ g_{\phi \phi} E^2 }{ \cal D }  + \mu^2 
            \right)/ \left( \frac{ 2 g_{t \phi} }{ \cal D } E 
	            \right) 
      }{ \frac{- {\cal D} }{ 2 g_{t \phi} E }  
       },
\end{equation}
which is a elliptic paraboloid and also {\it non-compact}. Therefore 
the value of the $p_i$ integrations are divergent.
Actually the surface such that ${g'}_{tt} = 0$ is the velocity of the
light surface (VLS). Beyond VLS (in region II) the co-moving 
observer must move  more rapidly  than the
velocity of light. 
Thus we will   assume  that the  system is in the region I.
For example, in  case of $\Omega_0 = 0$ the points 
satisfying ${g'}_{tt} =0$ are on the stationary limit 
surface. The region of the outside (inside) of the stationary 
limit surface corresponds to the region I (II).
In case of $\Omega_0 = \Omega_H$, where $\Omega_H$ is the 
angular velocity of the black hole, the region I corresponds 
to $r_H < r < r_{VLS}$, with $r_{VLS} \sim 1/\Omega_H $ 
approximately (for Kerr-Newman black hole).
VLS is an open, roughly, cylindrical surface \cite{MinHo}.

After some calculation  we obtain
\begin{eqnarray}
\nonumber
& & \beta W    \\
\nonumber
&=& \int_{region~ I} 
dr d \theta d \phi \int_{ \mu \sqrt{ - g_{tt}^{'} }}^\infty
	 \frac{d E}{2 \pi^2}  \frac{ \sqrt{g_4}}{( - g_{tt}^{'} )^2 }
	 E \left[ 
		  E^2 - ( - g_{tt}^{'} ) \mu^2  \right]^{1/2}
		  \ln \left(
	 1 - e^{- \beta (E - q \Gamma )}
	        \right)    \\
       &=&  - \beta
	 \int_{region~ I} 
	 dr d \theta d \phi \int_{ \mu \sqrt{ - g_{tt}^{'} }}^\infty
	 \frac{d E}{6 \pi^2}  \frac{ \sqrt{g_4}}{( - g_{tt}^{'} )^2 }
   \frac{  \left[ E^2 - ( - g_{tt}^{'} ) \mu^2  \right]^{3/2}}{
   e^{ \beta \left(
	   E - q \Gamma  \right) } - 1   },
	    \label{thpotential}
\end{eqnarray}
where we have integrated by parts. 
In particular when $\Omega_0 = a = e = q= 0$, 
the expression (\ref{thpotential}) 
coincides with the expression obtained by 't Hooft \cite{tHooft} and 
it is proportional to the volume of the optical space \cite{optical}.
In case of $q=0$ it reduces to the result in Ref. \cite{MinHo}.
It is easy to see that the  integrand diverges  as  
$h \rightarrow 0$.

Let $\mu = 0$. For a massless  charged scalar field 
the thermodynamic  potential reduces to\cite{Table} 
\begin{eqnarray}
\nonumber
   W       &= & -  \int_{region ~ I} 
	 dr d \theta d \phi \int_0^\infty
	 \frac{d E}{6 \pi^2}  \frac{ \sqrt{g_4}}{( - g_{tt}^{'} )^2 }
   \frac{   E^3 }{ e^{ \beta \left(
	   E - q \Gamma   \right) } - 1   }              \\
  \nonumber
    &= & -  \int_{region~  I} dr d \theta d \phi 
	 \frac{1}{ \pi^2 \beta^4}  
	 \frac{ \sqrt{g_4}}{( - g_{tt}^{'} )^2 }
	 \sum_{k = 1}^\infty 
	 \left( \frac{e^{k q \beta \Gamma}}{k^4}  \right)   \\
   &= & -  \int_{region  I} dr d \theta d \phi 
	 \frac{ \sqrt{g_4}}{ \pi^2 \beta^4_{local} }
	 \sum_{k = 1}^\infty 
	 \left( \frac{e^{k q \beta \Gamma}}{k^4}  \right),
\end{eqnarray}
where $\beta_{local} = \sqrt{ - g^{'}_{tt } } \beta$ is 
the reciprocal of the local Tolman temperature \cite{Landau}.
This form  is just the thermodynamic potential  
of a gas of massless  particles 
with chemical potential $q \Gamma$ at local
temperature $1/\beta_{local}$.
( We assumed  $ q \Gamma <0$. Then for  the antiparticle
 there is superradiance. Thus to obtain a finite value   
  ${\cal E} - m \Omega_0 - q \Phi_0 $ must great than  0. 
  See Eq. (\ref{po}).)

Now let us assume that $\Omega_0 = \Omega_H = \frac{a}{r_H^2 + a^2} $
and $ \Phi_0 = \Phi_H = 
\frac{ e r_H}{r_H^2 + a^2} $. In this case $\Gamma (r = r_H)  = 0$. 
Thus the leading behavior of  the thermodynamic potential for very 
small $h$ is given by 
\begin{eqnarray}
\nonumber
\beta W & \approx& -  \frac{\pi^2}{90 \beta^3} 
\int d \phi d \theta \int_{r_H + h}^L dr
 \sqrt{g_4} 
 \left\{
 \left( \frac{ g_{\phi \phi} }{ - {\cal D} }  \right)^2   
+ 2 \left( - \frac{g_{t \phi}}{g_{\phi \phi}}  - \Omega_H \right)^2 
\frac{g^4_{\phi \phi} }{(- {\cal D})^3}   
\right\} \\
\nonumber
 &\approx& - \frac{\pi^2}{90 \beta^3} \frac{A_H }{8 \kappa^3}
  \left\{
   2 \kappa \left(
   \frac{r_H}{a} + \frac{a}{r_H}
   \right) \tan^{-1} \left( \frac{a}{r_H} \right) \cdot 
   \frac{1}{h} + 
   \left(
      \left[
	-\frac{6 \kappa}{r_H} + \frac{2}{r_H^2 + a^2 }
	- \frac{3}{r_H^2} 
       \right]
     \right. \right. \\
     \nonumber
& &+~ \left. \left.
     \left[
     - \frac{6 a \kappa}{r_H^2} -\frac{6 \kappa}{a}
     + \frac{3}{a r_H} - \frac{3 a}{r_H^3} 
     \right] \tan^{-1} \left( \frac{a}{r_H} \right) 
     \right) \cdot \ln (h) + \cdot \cdot \cdot 
  \right\}  \\
&\equiv&  - \frac{\pi^2}{90 \beta^3 }  \frac{A_H}{8 \kappa^3}
\left\{ C \cdot \frac{1}{h} + D \cdot \ln(h) + \cdot \cdot \cdot
  \right\}, 
 \end{eqnarray}
 where $r_- = M - \sqrt{ M^2 - a^2 -e^2} $ 
 or in terms of the proper distance cut-off $\epsilon$
 \begin{eqnarray}
 \nonumber
\beta W  &\approx& -  \frac{\pi^2}{90 \beta^3} 
\int_{r = r_H} d \phi d \theta 
\sqrt{g_{\theta \theta} g_{\phi \phi}} 
 \int_{r_H + h}^L dr \sqrt{g_{rr}}
\left( \frac{g_{\phi \phi}}{g^2_{t \phi} -g_{tt}g_{\phi \phi}}
\right)^{3/2}   \\
&\approx& -  \frac{\pi^2}{180 (  \kappa \beta)^3 } 
\frac{A_H}{\epsilon^2}.
              \label{free2}
\end{eqnarray}
Here  $A_H$ is the area of the event horizon, and $\kappa$ is the 
surface gravity of the black hole,  and $\epsilon$ is the 
proper distance from the horizon to $r_H + h$:
\begin{eqnarray}
A_H &=&  \int_{r = r_H} \sqrt{g_{\theta \theta} g_{\phi \phi}} 
	d \theta d \phi = 4 \pi ( r_H^2  +a^2), \\
\kappa &=&  \frac{1}{2} \frac{r_H - r_-}{( r_H^2 + a^2) } 
 =  \frac{ ( M^2 - a^2 - e^2)^{1/2}}{
        2 M [ M + (M^2 - a^2 -e^2 )^{1/2}] - e^2 }, \\
\epsilon  &=& \int_{r_H}^{r_H +h}  \sqrt{g_{rr}} dr
\approx  2  \left( \frac{ r_H^2 + a^2 \cos^2 \theta}{2 r_H 
- 2 M} \right)^{1/2} \sqrt{h}.
\end{eqnarray}
The leading behavior of the entropy $S$ including the contribution 
of the antiparticle is 
\begin{eqnarray}
\nonumber
S &=& \beta^2 \frac{\partial}{\partial \beta} W   \\
 &\approx&     \frac{\pi^2}{90 \kappa^3 \beta^3} A_H 
 \left\{ 
   C \cdot \frac{1}{h} + D \cdot \ln(h) 
 \right\}
\end{eqnarray}
 or
\begin{equation} 
S   \approx  8  
\frac{\pi^2}{ 180 ( \kappa \beta)^3 }  \frac{A_H}{\epsilon^2}, 
\label{Entropy} 
\end{equation}
which   diverges as $h \rightarrow 0$. 
The divergences arise from the fact that  
the density of states for a given $E$
 diverges as $h$ goes to zero.

If we take $T$ as the Hartle-Hawking temperature 
$T_H = \frac{ \kappa}{2 \pi}$ 
( In this case the quantum 
state is the Hartle-Hawking vacuum state $|H \rangle$ \cite{Hartle}.) 
the  entropy  becomes
\begin{equation}
 S \approx 4  N A_H 
 \left\{
   C \cdot \frac{1}{h} + D \cdot \ln(h) 
 \right\}
\end{equation}
or
\begin{equation}
S \approx  N \frac{ A_H}{ \epsilon^2}   \label{result},  
\end{equation}
where $N$ is a  constant.
The entropy of a complex scalar field diverges quadratically  
in $\epsilon$ as the system approaches  the horizon,
or it diverges linearly and logarithmically in  $h$.
In case of  $ a= 0= q$ our result (\ref{result}) agrees with the   
result calculated by 't Hooft \cite{tHooft} 
and with one in Ref. \cite{ohta}.
These facts imply that the leading behavior of entropy  (\ref{result})
is  a general form.

As a  summary,  we showed the leading behavior of the entropy of the 
complex scalar field in Hartle-Hawking vacuum is proportional
to the horizon area, but diverges as the cut-off $h$ goes to zero. 
The  reason of the divergence is that at  the horizon
the density of states for a given $E$ diverges.
The particular points are followings:

1) There is a logarithmic divergence in the coordinate cut-off $h$.

2) The proper distance  cut-off $\epsilon $ 
 is  dependent on the coordinate $\theta$, which came from the 
axsisymmetric properties of the spacetime.

3) To obtain the entropy   proportional to the horizon area,
the angular velocity should  be equal to 
$\Omega_H$ and the chemical potential
should be equal to  $\Phi_H$.

4) In case of the extremal rotating black hole $(r_H = r_-)$
we can see that the cubic and the quadratic divergences in $h$  appear.
However  we must  consider only the case $a \leq 1/2 M$ to 
obtain a finite value \cite{MinHo}.

\begin{flushleft}
{\bf Acknowledgment}
\end{flushleft}

This work is partially supported by Korea Science and Engineering 
Foundation.

\end{document}